\documentclass{article}
\usepackage[T1]{fontenc} 
\usepackage[utf8]{inputenc} 
\usepackage{ismir,amsmath,cite,url}
\usepackage{graphicx}
\usepackage{color}
\usepackage{multirow}
\usepackage[normalem]{ulem}
\useunder{\uline}{\ul}{}
\usepackage[table,xcdraw]{xcolor}
\usepackage{float}

\title{The emotions that we perceive in music: the influence of language and lyrics comprehension on agreement}

\multauthor
{Juan Sebastián Gómez Cañón$^1$ \hspace{1cm} Perfecto Herrera $^1$} { \bfseries{ Emilia Gómez$^1$ \hspace{1cm} Estefanía Cano$^2$}\\
  $^1$ Music Technology Group, Universitat Pompeu Fabra, Spain\\
$^2$ Social and Cognitive Computing Department, A*STAR, Singapore\\
{\tt\small juansebastian.gomez@upf.edu}
}

\begin{document}

\maketitle
\begin{abstract}
In the present study, we address the relationship between the emotions perceived in pop and rock music (mainly in Euro-American styles with English lyrics) and the language spoken by the listener. Our goal is to understand the influence of lyrics comprehension on the perception of emotions and use this information to improve Music Emotion Recognition (MER) models. Two main research questions are addressed:
\begin{enumerate}
    \item Are there differences and similarities between the emotions perceived in pop/rock music by listeners raised with different mother tongues?
    \item Do personal characteristics have an influence on the perceived emotions for listeners of a given language?
\end{enumerate}

Personal characteristics include the listeners' general demographics, familiarity and preference for the fragments, and music sophistication \cite{Mullensiefen2014}.
Our hypothesis is that inter-rater agreement (as defined by Krippendorff's alpha coefficient) from subjects is directly influenced by the comprehension of lyrics. 
\end{abstract}
\section{Introduction}\label{sec:introduction}

Computational models able to predict emotions in audio signals are designed to evaluate emotionally-relevant acoustic features from a particular data set. These features are used to narrow down the so-called "semantic gap" from the physical properties of an audio signal (e.g., spectral flux, zero crossing rate) and the semantic concepts related to these properties (e.g., emotions, mood) \cite{Herrera2018, Celma2006}. The field of MER has been evaluated consistently since 2007 in the Music Information Retrieval Evaluation eXchange (MIREX) Audio Mood Classification task. This task consists of the classification of audio into five mood categories or clusters, containing different emotional tags. However, a "glass ceiling" of performance of these algorithms has been reached given the nature of the data and the annotations \cite{Celma2006}. This is due to the limited/low agreement in the annotations of the data sets and the generalized confusion between annotating the perceived/expressed emotion and the felt/induced emotion by music. The low agreement problem extends to other research problems in Music Information Research/Retrieval (MIR): music auto-tagging \cite{Bigand2013}, music genre recognition \cite{Sturm2013, Sturm2014}, and music similarity \cite{Flexer2016}.

Concretely, algorithms in MIR have natural upper bounds in performance due to the quality of human annotations, including the lack of an agreed methodology for gathering these annotations and the degree of subjectivity in the annotation process. Given this subjective nature of the perception of emotions, low agreement will lead to limited performance and poor generalization to new data. 

\section{Related work}\label{sec:related_work}

In this study we focus on the \textit{perception} of emotions in music, which requires a brief conceptualization of the logical progression in which a "musical communication" is achieved from the musician to the listener. This process begins with the \textit{expression} of an emotion by the performer or composer, which can be \textit{perceived} by a listener and perhaps \textit{arouse} a particular emotion. The listener can then regard the music as \textit{aesthetically valuable} and finally \textit{like} it \cite{Juslin2019}. However, this single-channel perspective is debatable, since perception and induction of emotions can operate in parallel (i.e., happy music can induce melancholy or sad music can induce excitement). \textit{Perception} refers to the emotion that a listener identifies when listening to music, which can be different to the emotion the musician intended to express/convey and the emotion aroused/induced. The reason to focus on the \textit{perception} of emotions is mainly twofold: it is relatively less influenced by situational factors of listening (environment, mood, etc.) \cite{Yang2011}, and listeners are generally consistent in their ratings of the emotional expression of music \cite{Juslin2019, Juslin2010Handbook, Juslin1997}. Additionally, certain musical features can be associated to particular emotions (e.g., \textit{happiness}: fast mean tempo, small tempo variability, bright timbre, sharp duration contrasts, \textit{sadness}: slow mean tempo, legato articulation, low sound level, dull timbre, slow vibrato, \textit{fear}: staccato articulation, very low sound level, large sound level variability, large timing variations, etc.). The relation between features and the expression of emotion has been widely researched in the literature \cite{Juslin2019, Juslin2010Handbook}.

Certain attributes and characteristics of emotion are important to continue formalizing the definition of emotions: 1) emotions appear to vary their intensity (e.g., irritation - rage), 2) emotions appear to involve distinct qualitative feelings (e.g., feeling of being afraid - nostalgic), 3) some pairs of emotions appear more similar than others (e.g., joy and contentment versus joy and disgust, and 4) some emotions appear to have opposites (e.g., happiness - sadness, love - hate, calm - worry) \cite{Juslin2010Handbook}.  These attributes have led to the definition of two different dominant views or approaches for emotion representation \cite{Zentner2008}:
\begin{itemize}
    \item \textit{Categorical approach}: emotions are represented as categories, distinct from each other - such as happiness, sadness, anger, surprise, and fear. Hevner wrote a seminal paper on finding and grouping 66 adjectives into 8 groups of emotions \cite{Hevner1936}. Ekman defined a set of basic emotions in terms of human goal-relevant events that occurred during evolution \cite{Ekman1992}. The discrete categories of emotions can also be regarded as colors: there are different shades of sadness, but there is an abrupt change from one category (sad - blue) to another (angry - red) \cite{Juslin2019}. In this case, major drawbacks are: 1) the number of primary emotion categories results too small compared to the richness of music emotion perceived by humans (i.e., poor resolution) and 2) the high ambiguity of using language to describe rich human emotions \cite{Yang2011}.
    \item \textit{Dimensional approach}: emotions are conceptualized based on their positions on a small number of dimensions; mainly valence and arousal. Russell popularized the two-dimensional circumplex model, where the valence dimension describes the pleasantness or positiveness of the emotion and arousal describes the activation or energy (i.e., happiness would have positive arousal and valence) \cite{Russell1980}. However, the major drawback is that categories are not mutually exclusive and tend to overlap (i.e., rage-anger), making the mapping of categories on the dimensional space vague and unreliable \cite{Yang2011}.
\end{itemize}

In theory, certain categories of emotions could be directly mapped from a categorical to a dimensional approach \cite{Juslin2019}. However this mapping is not necessarily straightforward, when trying to map emotions such as \textit{nostalgia} or \textit{transcendence}. Although both approaches offer several variations and advantages over each other, some researchers in the field of psychology have chosen categorical models over dimensional models \cite{Juslin2019}. The main reasons for this are: 1) categories are needed to capture and do justice to complex human emotions experienced and perceived by listeners (i.e., complex emotions such as nostalgia, awe, and transcendence cannot be reduced to a deterministic value on VA dimensional space), and 2) emotions tend to show discreteness and boundaries between them (e.g., content - happy), rather than continuity when tested \cite{Laukka2005}.

In the particular case of instrumental classical music, researchers have attempted to find the relationship between listeners' characteristics and perceived emotions \cite{Schedl2018}. Web surveys were designed to collect information about personal characteristics and the perception of 15 segments of 3rd Symphony \textit{Eroica} by L.V. Beethoven (classical instrumental music). They used a selection of the GEMS categorical model of emotions \cite{Zentner2008}, including the following emotions: transcendence, peacefulness, power, joyful activation, tension, sadness, anger, disgust, fear, surprise, and tenderness. With respect to personal characteristics, the results suggest that: 1) the perception of transcendence and power correlate significantly with basic user characteristics; 2) participants trained on classical music tend to disagree more on perceived emotions of peacefulness, tension, sadness, anger, disgust, and fear; 3) the agreement among perceived emotions decreases with increasing familiarity with the piece. With respect to the correlation of perceived music characteristics irrespective of listener characteristics, they have found that: 1) there are substantial correlations between pairs of anger, fear, and disgust; 2) peacefulness is negatively correlated with power and tension, but positively correlated with tenderness; 3) power is significantly correlated with tension and anger; 4) transcendence and surprise do not show significant correlation with other aspects.

The goal of this work is to study the relationship between listeners' demographics preference, familiarity, musical knowledge, and native language with agreement of perceived emotions in music. We aim to characterize perceived emotion with respect to these factors, and attempt to replicate previous studies that show lower agreement in perceived emotions among subjects with more musical experience and knowledge. In order to achieve this, we analyze agreement of emotion ratings of musical fragments from different styles that have been previously annotated. The rest of the paper is structured as follows: in Section 3 we detail the methodology of our study, including the selected measures, musical excerpts and annotation gathering scheme. Section 4 later provides partial results of our study which are later discussed in Section 5.

\section{Methodology}\label{sec:methodology}

\subsection{Agreement metrics}

Following \cite{Schedl2018}, we use inter-rater reliability statistics (i.e., Krippendorff's $\alpha$) to assess the agreement of the annotated data with respect to different individual characteristics and the understanding of the semantic content of music \cite{Krippendorff2004}. Krippendorff's coefficient is defined as: "$\alpha$ is the extent to which the proportion of the differences (amongst all observations) that are in the error deviates from perfect agreement. [...] It is the proportion of the observed to expected above-chance agreement". In general, $\alpha$ is defined as:
\begin{equation}
    \alpha = 1 - \frac{D_o}{D_e}
\end{equation}
where $D_o$ is the measure of observed disagreement:
\begin{equation}
    D_o = \frac{1}{n} \sum_{c} \sum_{k} o_{ck} \cdot {}_{metric}\delta_{ck}^{2}
\end{equation}
and $D_e$ is a measure of the expected disagreement given chance:
\begin{equation}
    D_e = \frac{1}{n(n-1)} \sum_{c} \sum_{k} n_c \cdot n_k \cdot {}_{metric}\delta_{ck}^{2}
\end{equation}
The variables $o_{ck}$, $n_c$, $n_k$  are the frequencies of values of observed coincidences of $c$ and $k$ values or ranks and $n$ is the total amount of paired $c-k$ values or ranks. In this case if $c = k$, these are the observed coincidences for these values (i.e., all users have the same rating on a particular emotion). On the other hand, if $c {\ne} k$ there are mismatched ratings. Advantages of using $\alpha$ are that it is suitable for any number of observers, any type of metric (nominal, interval, ordinal), it can handle incomplete or missing data, and does not require a minimum of sample size. When disagreement is absent ($D_o$ = 0), there is perfect reliability ($\alpha$ = 1). Conversely, when agreement and disagreement are a matter of chance ($D_e$ = $D_o$), there is absence of reliability ($\alpha$ = 0). Nevertheless, $\alpha$ could be smaller than zero due to sampling errors (too small sample sizes) and systematic disagreements (agreement below what would be expected by chance). 

Depending on the data,  ${}_{metric}\delta_{ck}$ is the difference function, the squared difference between any two values or ranks $c$ and $k$, function depending on the type of metric. In the case of ordinal metrics (i.e., using Likert response formats for emotional ratings) standardized as $0 \leq {}_{ordinal}\delta_{ck}^{2} \leq 1$:
\begin{equation}
    {}_{ordinal}\delta_{ck}^{2} = \left( {\frac{\sum_{g=c}^{g=k} n_g - \frac{n_c + n_k}{2}}{n - \frac{n_{c_{max}} + n_{c_{min}}}{2}}} \right) ^2
\end{equation}
where $c_{max}$ is the largest and $c_{min}$ is the smallest rank among all ranks.

\subsection{Music material}

For our agreement study, we selected a set of 22 music fragments from the 4Q emotion data set \cite{Panda2018} which has been previously annotated with categories in the four arousal-valence quadrants: Q1 (A+V+), Q2 (A+V-), Q3 (A-V-), Q4 (A-V+). We use fragments of pop and rock music since these musical styles can be considered as neutral and homogeneous even when sung in different languages. The music fragments (with 30 seconds duration) were collected from AllMusic API, and 289 emotion tags were selected from the original AllMusic Tags and intersected with the Warriner’s list \cite{Warriner2013}. In this way, emotion tags were mapped to AV space. Finally, they conducted a manual blind validation to remove inadequate fragments (e.g., containing noise or speech). The data set is balanced with 225 fragments per quadrant and a total of 900 clips.

We use the emotion tags of the Geneva Emotion Music Scale (GEMS) \cite{Scherer1995} and a subset from basic emotions \cite{Ekman1992} to rate the different fragments (see Table \ref{tab:emotions_syn}). In comparison to \cite{Schedl2018}, we replaced \textit{Disgust} by  \textit{Bitterness} attempting to improve the balance of the number of emotions per quadrant (see Table \ref{tab:emotions_syn}). Nonetheless, Q3 only contains two emotions, whereas other quadrants contain three. To perform the selection of fragments for this experiment, we performed a query for the selected emotions. Since some emotions were not found in the metadata, synonyms were used to select the songs, as seen in Column 3. After the automatic selection of songs by the query, we manually selected two fragments with lyrics per emotion (only two songs are instrumental).

\begin{table}[ht!]\label{tab:emotions_syn}
\centering
{\scriptsize
\begin{tabular}{|l|l|l|}
\hline
\multicolumn{1}{|c|}{\textbf{Quadrants}} & \multicolumn{1}{c|}{\textbf{Emotions}} & \multicolumn{1}{c|}{\textbf{Synonyms}} \\ \hline
\multirow{3}{*}{Q1 (A+V+)} & Joyful activation & joy \\ \cline{2-3} 
 & Power & - \\ \cline{2-3} 
 & Surprise & - \\ \hline
\multirow{3}{*}{Q2 (A+V-)} & Anger & angry \\ \cline{2-3} 
 & Fear & anguished \\ \cline{2-3} 
 & Tension & tense \\ \hline
\multirow{2}{*}{Q3 (A-V-)} & Bitterness & bitter \\ \cline{2-3} 
 & Sadness & sad \\ \hline
\multirow{3}{*}{Q4 (A-V+)} & Tenderness & gentle \\ \cline{2-3} 
 & Peace & - \\ \cline{2-3} 
 & Transcendence & spiritual \\ \hline
\end{tabular}
}
\caption{Selected emotions and query synonyms for song selection.}
\end{table}

In the case of music emotion studies, very few studies have explored different styles of music. Also several studies refer to the WEIRDness of music psychology studies\cite{Henrich2010} as the fact that the participants from experiments usually come from \textbf{W}estern, \textbf{E}ducated, \textbf{I}ndustrialized, \textbf{R}ich, and \textbf{D}emocratic countries. Since we base our approach on re-annotating previously annotated data, it is difficult to find annotated emotion data sets that contain music with lyrics in other languages than English. Nonetheless, we managed to include 3 songs in Spanish. The summary of previously annotated emotions, quadrants, language of lyrics and song information can be seen in Table \ref{tab:summary_songs}.

\begin{table}[ht!]
{
\scriptsize
\centering
\begin{tabular}{|l|l|l|l|}
\hline
\multicolumn{1}{|c|}{\textbf{Emotion}} & \multicolumn{1}{c|}{\textbf{Q}} & \multicolumn{1}{c|}{\textbf{Lang.}} & \multicolumn{1}{c|}{\textbf{Artist - Song}} \\ \hline
\multirow{2}{*}{Anger} & \multirow{2}{*}{Q2} & Eng. & Disincarnate - In Sufferance \\ \cline{3-4} 
 &  & Inst. & Obituary - Redneck Stomp \\ \hline
\multirow{2}{*}{Bitterness} & \multirow{2}{*}{Q3} & Eng. & Liz Phair - Divorce Song \\ \cline{3-4} 
 &  & Eng. & Lou Reed - Heroine \\ \hline
\multirow{2}{*}{Fear} & \multirow{2}{*}{Q2} & Inst. & Joe Henry - Nico Lost One Small Buddha \\ \cline{3-4} 
 &  & Eng. & Silverstein - Worlds Apart \\ \hline
\multirow{2}{*}{Joy} & \multirow{2}{*}{Q1} & Eng. & Taio Cruz - Dynamite \\ \cline{3-4} 
 &  & Eng. & Miami Sound Machine - Conga \\ \hline
\multirow{2}{*}{Peace} & \multirow{2}{*}{Q4} & Eng. & Jim Brickman - Simple Things \\ \cline{3-4} 
 &  & Spa. & Gloria Estefan - Mi Buen Amor \\ \hline
\multirow{2}{*}{Power} & \multirow{2}{*}{Q1} & Eng. & Ultra Montanes - Anyway \\ \cline{3-4} 
 &  & Eng. & Rose Tattoo - Rock n Roll Outlaw \\ \hline
\multirow{2}{*}{Sadness} & \multirow{2}{*}{Q3} & Eng. & Motorhead - Dead and Gone \\ \cline{3-4} 
 &  & Spa. & Juan Luis Guerra - Sobremesa \\ \hline
\multirow{2}{*}{Surprise} & \multirow{2}{*}{Q1} & Eng. & The Jordanaires - Hound Dog \\ \cline{3-4} 
 &  & Eng. & Shakira - Animal City \\ \hline
\multirow{2}{*}{Tenderness} & \multirow{2}{*}{Q4} & Eng. & Celine Dion - Beautiful Boy \\ \cline{3-4} 
 &  & Spa. & Beyonce - Amor Gitano \\ \hline
\multirow{2}{*}{Tension} & \multirow{2}{*}{Q2} & Eng. & Pennywise - Pennywise \\ \cline{3-4} 
 &  & Eng. & Squeeze - Here Comes That Feeling \\ \hline
\multirow{2}{*}{Transc.} & \multirow{2}{*}{Q4} & Eng. & Steven C. Chapman - Made for Worshipping \\ \cline{3-4} 
 &  & Eng. & Matisyahu - On Nature \\ \hline
\end{tabular}
}
\caption{Song selection with emotion and quadrant information.}
\label{tab:summary_songs}
\end{table}

\subsection{Annotation Methodology}

\begin{figure*}[h!]
  \includegraphics[width=\linewidth]{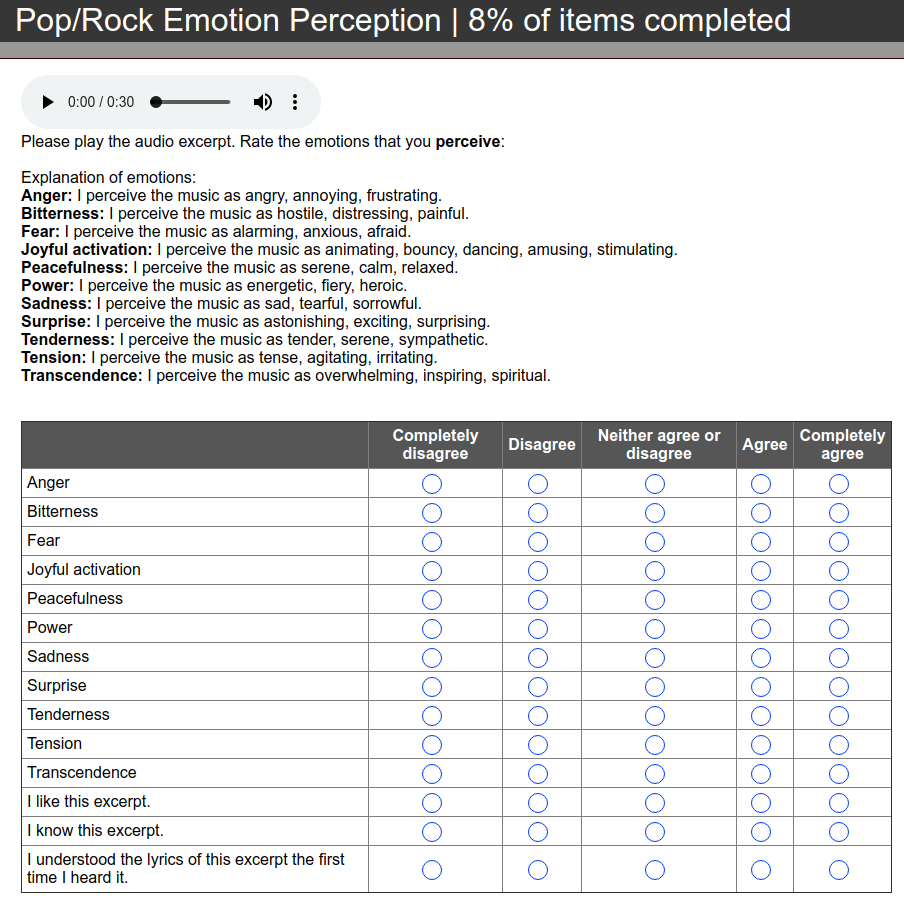}
  \caption{Example of emotion rating survey in English.}
  \label{fig:survey}
\end{figure*}

\begin{table*}[!ht]
{\scriptsize
\begin{tabular}{lll|l|l|l|l|l|l|l|l|l|l|l|}
\cline{4-14}
 &  &  & \multicolumn{3}{c|}{\textbf{Q1}} & \multicolumn{3}{c|}{\textbf{Q2}} & \multicolumn{2}{c|}{\textbf{Q3}} & \multicolumn{3}{c|}{\textbf{Q4}} \\ \hline
\multicolumn{1}{|l|}{\textbf{Configuration}} & \multicolumn{1}{l|}{\textbf{Ratings}} & \textbf{\%} & \textbf{joy} & \textbf{surp.} & \textbf{pow.} & \textbf{ang.} & \textbf{fear} & \textbf{tens.} & \textbf{sad} & \textbf{bit.} & \textbf{peace} & \textbf{tend.} & \textbf{trans.} \\ \hline
\rowcolor[HTML]{FFFFC7} 
\multicolumn{1}{|l|}{\cellcolor[HTML]{FFFFC7}All} & \multicolumn{1}{l|}{\cellcolor[HTML]{FFFFC7}23562/23562} & 100.00\% & 0.401 & 0.064 & 0.268 & 0.355 & 0.193 & 0.274 & 0.286 & 0.24 & 0.367 & 0.346 & 0.057 \\ \hline
\multicolumn{1}{|l|}{By Preference (>3)} & \multicolumn{1}{l|}{9427/23562} & 40.01\% & 0.407 & 0.075 & 0.267 & \cellcolor[HTML]{FFCCC9}0.282 & 0.165 & \cellcolor[HTML]{FFCCC9}0.185 & 0.32 & 0.228 & 0.368 & 0.35 & 0.064 \\ \hline
\multicolumn{1}{|l|}{By Preference (<3)} & \multicolumn{1}{l|}{7755/23562} & 32.91\% & \cellcolor[HTML]{FFCCC9}0.321 & 0.039 & 0.261 & 0.384 & 0.179 & \cellcolor[HTML]{32CB00}0.332 & 0.251 & 0.203 & 0.372 & 0.348 & 0.046 \\ \hline
\multicolumn{1}{|l|}{By Familiarity (>3)} & \multicolumn{1}{l|}{3795/23562} & 16.11\% & \cellcolor[HTML]{32CB00}0.456 & 0.052 & \cellcolor[HTML]{FFCCC9}0.153 & 0.332 & \cellcolor[HTML]{32CB00}0.262 & \cellcolor[HTML]{FFCCC9}0.157 & 0.329 & \cellcolor[HTML]{32CB00}0.318 & \cellcolor[HTML]{FFCCC9}0.224 & \cellcolor[HTML]{FFCCC9}0.194 & 0.034 \\ \hline
\multicolumn{1}{|l|}{By Familiarity (<3)} & \multicolumn{1}{l|}{18183/23562} & 77.17\% & \cellcolor[HTML]{FFCCC9}0.313 & 0.047 & 0.277 & 0.345 & 0.167 & 0.292 & 0.243 & 0.194 & 0.408 & 0.379 & 0.069 \\ \hline
\multicolumn{1}{|l|}{By Understanding (>3)} & \multicolumn{1}{l|}{11616/23562} & 49.30\% & 0.424 & 0.079 & 0.268 & \cellcolor[HTML]{FFCCC9}0.28 & 0.183 & \cellcolor[HTML]{FFCCC9}0.21 & 0.324 & 0.274 & 0.334 & 0.336 & 0.043 \\ \hline
\multicolumn{1}{|l|}{By Understanding (<3)} & \multicolumn{1}{l|}{8261/23562} & 35.06\% & \cellcolor[HTML]{FFCCC9}0.336 & 0.035 & 0.251 & 0.375 & 0.181 & 0.318 & \cellcolor[HTML]{FFCCC9}0.231 & \cellcolor[HTML]{FFCCC9}0.181 & 0.362 & 0.299 & 0.067 \\ \hline
\end{tabular}
}
\caption{Krippendorff's $\alpha$ for each emotion for all participants filtered by preference, familiarity, and lyrics comprehension (positive and negative). We use a 5-point Likert response format, we consider positive ratings as higher than 3 (neither agree or disagree) and negative as less than 3}
\label{tab:filters_agree}
\end{table*}
Our study is structured as follows: we first provide a brief explanation to show participants the difference between \textit{induced} and \textit{perceived} emotions. Additionally, we use synonyms for each emotion to clarify the annotations (see Figure \ref{fig:survey}). To collect user ratings, we created online surveys in four languages (Spanish\footnote{\url{https://www.psytoolkit.org/cgi-bin/psy2.5.3/survey?s=pa92w}}, English\footnote{\url{https://www.psytoolkit.org/cgi-bin/psy2.5.3/survey?s=hVTWu}}, German\footnote{\url{https://www.psytoolkit.org/cgi-bin/psy2.5.3/survey?s=gUEny}} and Mandarin\footnote{\url{https://www.psytoolkit.org/cgi-bin/psy2.5.3/survey?s=APaWq}}) using two excerpts per emotion, for a total of 22 excerpts. Besides emotion ratings, we also collect personal information about musical knowledge, musical taste, listeners' familiarity with the stimuli, listeners' understanding of the lyrics, and demographics. 

All audio excerpts were normalized from -1 to 1 and every participant must complete a previous step to set the volume of the survey with respect to a 1 KHz sinusoid. Figure \ref{fig:survey} shows an example of one of the questions from the survey. In order to measure the musical knowledge from the participants, we use the Music Sophistication Index\cite{Mullensiefen2014} and make the results available to the participants at the end of the survey as a small thank you.

\section{Results and Discussion}\label{sec:evaluation}

The participation to this moment has been unbalanced, as Tables \ref{tab:filters_agree} and \ref{tab:language_agree} show. The color coding shows \textit{green} when the agreement of the emotion is higher by 0.05 than the agreement measured across 126 participants from all languages, which are coded in \textit{yellow}. Conversely, the cell is \textit{red} when the difference is less than -0.05. Initially, it is possible to see that the agreement over complex emotions is very low (approximately 0.2), such as \textit{bitterness}, \textit{fear}, \textit{power}, \textit{surprise}, and \textit{transcendence}. On the other hand, a higher agreement is reached for more basic emotions, such as \textit{anger}, \textit{joy}, \textit{peace}, \textit{sadness}, and \textit{tenderness}. 

We use the word \textit{filter}, since we are removing the ratings of fragments that were or not preferred, familiar or understood by the listeners. Inter-rater agreement can also be evaluated across every song with positive and negative filters. Since we use a 5-point Likert response format, we consider positive ratings as higher than 3 (neither agree or disagree) and negative as less than 3 (see Figure \ref{fig:survey}). From now on we refer to the filters and emotions as a combination pair, i.e. \textit{joy - preference} means the positive and negative \textit{preference} for the emotion \textit{joy}.  In the case of Q1 and Q3, agreement is higher with positive filter than negative filtering for all the filters (with the exception of \textit{power - familiarity}. Quadrants Q1 (A+V+) and Q3 (A-V-) can be considered as very distinct and "universal". Conversely, quadrants Q2 and Q4 show the opposite behavior, agreement tends to be lower negative than positive filtering for all the filters (with the exceptions of \textit{fear - familiarity}, \textit{fear - understanding}, \textit{tenderness - preference}, \textit{transcendence - preference}, and \textit{tenderness - understanding}). We refer to exceptions when the results do not adjust to the tendencies described above.

\begin{table}[!htp]
{\scriptsize
\begin{tabular}{|l|l|l|l|l|
>{\columncolor[HTML]{FFFFC7}}l |}
\hline
\textbf{Emotions} & \textbf{Eng. (26)} & \textbf{Spa. (56)} & \textbf{Man. (27)} & \textbf{Ger. (17)} & \multicolumn{1}{c|}{\cellcolor[HTML]{FFFFC7}\textbf{All (126)}} \\ \hline
anger & \cellcolor[HTML]{32CB00}0.429 & \cellcolor[HTML]{FFCCC9}0.311 & 0.367 & \cellcolor[HTML]{32CB00}0.482 & 0.364 \\ \hline
bitter & \cellcolor[HTML]{32CB00}0.278 & 0.209 & 0.155 & \cellcolor[HTML]{32CB00}0.278 & 0.202 \\ \hline
fear & \cellcolor[HTML]{32CB00}0.241 & 0.175 & \cellcolor[HTML]{FFCCC9}0.091 & 0.207 & 0.171 \\ \hline
joy & \cellcolor[HTML]{FFCCC9}0.304 & \cellcolor[HTML]{32CB00}0.437 & \cellcolor[HTML]{FFCCC9}0.311 & \cellcolor[HTML]{32CB00}0.476 & 0.372 \\ \hline
peace & 0.401 & 0.332 & 0.401 & \cellcolor[HTML]{32CB00}0.438 & 0.371 \\ \hline
power & \cellcolor[HTML]{32CB00}0.379 & 0.287 & 0.296 & 0.325 & 0.289 \\ \hline
sad & 0.330 & 0.343 & 0.279 & \cellcolor[HTML]{32CB00}0.378 & 0.326 \\ \hline
surprise & 0.041 & 0.055 & 0.068 & \cellcolor[HTML]{32CB00}0.218 & 0.075 \\ \hline
tender & 0.444 & \cellcolor[HTML]{FFCCC9}0.314 & \cellcolor[HTML]{32CB00}0.452 & \cellcolor[HTML]{32CB00}0.581 & 0.396 \\ \hline
tension & 0.264 & 0.324 & 0.282 & 0.323 & 0.296 \\ \hline
transc. & 0.080 & 0.049 & 0.083 & \cellcolor[HTML]{FFCCC9}-0.012 & 0.057 \\ \hline
\end{tabular}
}
\caption{Krippendorff's $\alpha$ for each emotion and four questionnaires.}
\label{tab:language_agree}
\end{table}

Furthermore unbalanced participation in the surveys forces to evaluate all subjects simultaneously. It is important to note that Table \ref{tab:language_agree} evaluates agreement over subjects, while Table \ref{tab:filters_agree} evaluates agreement over all subjects \textit{that selected a song with a particular filter}. For this reason, Table \ref{tab:filters_agree} contains information about the number of ratings taken into account when using the filters.

It is important to acknowledge in the case of \textit{familiarity} filter, only 16\% of the ratings were evaluated as positive. This means that the selected music (and personal memories attached to them) should have a smaller influence on the experiment and that this filter is unbalanced while evaluating agreement. Since agreement is evaluated upon the number of subjects and number of ratings made, the comparison of the familiarity filter should be very noisy. The exceptions  \textit{fear - familiarity} and \textit{power - familiarity} show a difference of 12\% that could be explained by this. On the other hand, the remaining pairs of exceptions (\textit{fear - understanding}, \textit{tenderness - preference}, \textit{transcendence - preference}, and \textit{tenderness - understanding}) show a variability of 1-3\% when using the filters. The remaining 27 combinations of pairs show the overall behavior mentioned previously with a variation of 1-18\% of the tendencies depending on the emotion and filter. 

\section{Conclusions}

Our results have confirmed that basic emotions will have higher universal agreement, while complex ones will show the opposite (as seen in Table \ref{tab:language_agree}). However, agreement in our experiment appears to be lower than values reported in \cite{Aljanaki2017, Flexer2016}. Following \cite{Krippendorff2004}, data should only be considered reliable when obtaining $\alpha > 0.6$.

Our initial results suggest that lyrics comprehension (LC) improves agreement for emotions in quadrants Q1-Q3, and decreases it for quadrants Q2-Q3. This relates to the type of emotions that we find related to each quadrant and the subjectivity that has been previously researched regarding valence. This has given us new understanding of the effect of LC and its impact on different emotions: 1) in the case of Q1-Q3, better understanding could lead to similar emotions being recognized and result in a higher agreement in the intra-linguistic setting, 2) in the case of Q2-Q4, better understanding could lead to a finer criteria when judging and result in a lower agreement in the intra-linguistic setting. 

This study is part of an ongoing research which is now focusing on improving the present study.  We consider balancing the styles with respect to the different languages. This would mean to have music relating to each culture/language and all languages in the study should be considered. Additionally, it is highly debatable that pop and rock are in fact neutral and homogeneous since several variations across the world show different ways to convey emotions (e.g., Hindi pop). Lastly, the experiment could have biased the responses in the sense that asking for the understanding forces the subjects to pay attention to the lyrics, losing ecological validity. Future studies will consider other types of tests to prove if subjects effectively understood the lyrics (e.g., using a "fill in the lyrics" question after listening to each fragment).

\bibliography{ISMIRtemplate}

%
%
%
%

\end{document}